\documentclass[aps,prb,twocolumn,showpacs,floatfix]{revtex4}
\usepackage{dcolumn}
\usepackage{graphicx,amssymb,amsmath}
\usepackage{bm}
\usepackage{natbib}
\usepackage{epstopdf}
\usepackage{stmaryrd}
\usepackage{latexsym}
\usepackage{amsfonts}

\begin{document}

\title{Effects of disorder on atomic density waves and spin-singlet dimers
       in one-dimensional optical lattices}

\author{Gao Xianlong}
\email{gaoxl@zjnu.cn}
\affiliation{Department of Physics, Zhejiang Normal University, Jinhua, Zhejiang Province, 321004, China}

\date{\today}
\begin{abstract}
Using the Bethe-ansatz density-functional theory, we study a one-dimensional Hubbard model of confined attractively interacting fermions in the presence of a uniformly distributed disorder. The strongly-correlated Luther-Emery nature of the attractive one-dimensional Hubbard model is fully taken into account as the reference system in the density-functional theory. The effects of the disorder are investigated on the atomic-density waves in the weak-to-intermediate attractive interaction and on the spin-singlet dimers of doubly occupied sites in the strongly attractive regime. It is found that atomic density waves are sensitive to the disorder and the spin-singlet dimers of doubly occupied sites are quite unstable against the disorder. We also show that very weak disorder could smear the singularities in the stiffness, thus suppress the spin-singlet pairs.
\end{abstract}

\maketitle

\section{Introduction}
\label{sect:intro}
Disorder plays an important role in physical properties of quantum many-body systems, particularly in solid state physics. Examples of phenomena related to disorder include a drastic suppression of superfluidity of $^4$He in nanoporous media,~\cite{porus} an apparent metal-insulator quantum phase transition in a two-dimensional electron liquid induced by the interplay between long-range Coulomb interactions and disorder,~\cite{MIT} a strong dependence of supersolidity on the amount of disorder in solid $^4$He,~\cite{supersolid} and a halted expansion of a Bose-Einstein condensate (BEC) in a random potential.~\cite{coldrandom}

Intriguing achievements of ultracold atomic systems in optical lattices have opened an exciting field in manipulating many of the model Hamiltonian parameters in a clean and controlled way. For example, experimentalists can control the on-site interaction strength, confining potentials, the effective dimensionality, and different forms of disorder to simulate phenomena in solid-state physics like the superfluid-Mott-insulator transition~\cite{Greiner} and BEC-BCS crossover of long range phase coherence of fermionic pairs.~\cite{Chin} The on-site interaction can be tuned either indirectly by changing the strength of the lasers that create the optical lattice potential or directly by using a Feshbach resonance, which allows to change the scattering length from $-\infty$ to $\infty$. Optical lattices can be used to find new exotic quantum states of matter, for example, by trapping Bose-Fermi mixtures~\cite{Modugno} and multiple hyperfine states~\cite{Hofstetter} that have no direct analogues in condensed matter systems. Besides the clean realizations of many condensed-matter lattice models in optical lattices, disorder can also be generated in optical lattices.

Disorder in optical lattices is attracting increasing attention of many theorists and experimentalists due to its unique flexibility and unprecedented controllability available in producing and observing disordered quantum degenerate gases. The unique controllability gains the research on optical lattices of disorder much advantage over the disorder research in electronic systems, where the disorder is fixed by a specific realization of the sample and the interaction is deemed to be the long-range Coulomb one.~\cite{long-range coulomb} Disorder in optical lattices can be created in different ways: an unbounded speckle pattern produced by shining a laser beam through a diffusive plate,~\cite{coldrandom,speckle} bounded incommensurate bichromatic lattices produced by combining the primary optical lattice with a secondary lattice (not random but completely deterministic quasidisorder),~\cite{Roth,Lye,Lye-boseglass} and a superimposed disorder.~\cite{Schulte} Other novel kinds of disorder have been proposed, for example, a bounded disorder in the strength of interatomic interactions, which can be realized near a Feshbach resonance,~\cite{Gimperlein} and randomly located impurity atoms with a different internal state trapped at the nodes of an optical lattice.~\cite{Gavish} Several proposals have been put forward in regard to disordered ultracold bosonic atom gases, where a rich variety of exotic new quantum phases like the Bose glass, the Bose-Anderson glass, and the Mott glass have been predicted.~\cite{Schulte,Fisher,Damski} Anderson localization is predicted to be observable from the expansion of an initially confined interacting one-dimensional (1D) BEC in a weak disorder.~\cite{Palencia}

Cold Bose gases have been successfully trapped in 1D geometries. A Tonks-Girardeau gas of bosonic $^{87}$Rb atoms~\cite{Paredes} was realized experimentally in a 1D optical lattice. A $^{87}$Rb atomic gas inside a disordered 1D optical lattice (OL) has been used to study the interplay between repulsive interactions and disorder.~\cite{Schulte} Towards the experimental test of predicted exotic phases, the Florence group gave us the first experimental hint of Bose glass by measuring the excitation energy spectrum and coherence.~\cite{Lye-boseglass} In the case of fermionic atom gases, however, it is more difficult to cool them down because of their Pauli-limited collision rate, different from the s-wave collisional pattern of bosonic atoms. Up to now a two-component Fermi gas of $^{40}$K atoms has been prepared in a quasi-1D geometry.~\cite{moritz_2005} An interacting Fermi gas of $^{40}$K atoms has also been demonstrated in three-dimensional optical lattices.~\cite{Kohl} We think that trapping interacting gases of fermionic atoms in 1D optical lattices is within the reach of the present-day techniques.

These recent developments in experiment make an exploration of fermions in disordered optical lattices timely. Some proposals related to fermionic atom gases in disordered OLs have already been made to revisit disordered condensed matter models. For example, in the work of Paredes {\it et al.},~\cite{Paredes_superlattice} it was reported that fermionic atoms in optical superlattices exhibit strongly correlated phenomena, from Kondo singlet formation to magnetism of localized spins. Yamashita {\it et al.} studied the characteristic properties of fermions trapped in a 1D optical superlattice with 2-site periodicity and found that three different insulating phases (of band-, bond-charge-density-wave- and Mott-type) can emerge.~\cite{Yamashita} Repulsive interacting Fermi gases in disordered 1D OLs have been studied in Ref.~\onlinecite{gao_PRB_repulsive}, where the effects of disorder on local Mott-insulating and band-insulating regions have been analyzed in detail.

Motivated by these experimental and theoretical scenario, in this paper we study the interplay between attractive interactions and uniform randomness in a two-component Fermi gas loaded in a 1D OL under harmonic confinement. We investigate systematically the effects of a random potential on these systems making use of Bethe-ansatz density-functional theory.~\cite{soft,Gao_PRBlong} In the clean limit, the system exhibits atomic-density waves for weak-to-intermediate attractive interactions and band-insulating regions for strongly attractive interactions.~\cite{Gao_PRL}

The structure of the paper is the following. In Sec. II, the model Hamiltonian is briefly introduced. In Sec. III, we present a lattice version of density-functional theory based on the exactly solvable 1D homogeneous Hubbard model. Finally, numerical results and some conclusions are shown in Sec. IV.

\section{The 1D random attractive Fermi-Hubbard model}
\label{sect:model}
A 1D Fermi gas trapped by an external harmonic potential and by a stationary optical potential can be described by a 1D single-band Fermi-Hubbard model, if we assume that the lattice potential is deep enough and that the energy separation between the first and the second band of the lattice is much larger than other energy scales involved:
\begin{eqnarray}\label{eq:hubbard}
\hat {H}_s&=&-t\sum_{i=1,\sigma}^{N_s}({\hat c}^{\dagger}_{i\sigma}{\hat c}_{i+1\sigma}+{\rm H}.{\rm c}.)+
U\sum_{i=1}^{N_s}\,{\hat n}_{i\uparrow}{\hat n}_{i\downarrow}
\nonumber\\
&&+V_2\sum_{i=1}^{N_s}(i-N_{s}/2)^2{\hat n}_i\,.
\end{eqnarray}
Here $\sigma=\uparrow,\downarrow$ is a pseudospin-$1/2$ label for two internal hyperfine states, ${\hat n}_i= \sum_\sigma {\hat n}_{i\sigma}=\sum_\sigma {\hat c}^{\dagger}_{i\sigma}{\hat c}_{i\sigma}$ is the total site occupation operator, $t$ is the tunneling between nearest neighbors, $U$ is the on-site attractive interaction and $V_2$ is the strength of harmonic potential. A system of $N_s$ lattice sites and $N_f$ interacting fermions is considered. The 1D random attractive Fermi-Hubbard model is written as,
\begin{equation}\label{eq:hubbard-disorder}
\hat {H}=\hat {H}_s+\sum_{i=1}^{N_s}\varepsilon_i\,{\hat n}_i\,.
\end{equation}
The effect of disorder is simulated by the last term in Eq.~(\ref{eq:hubbard-disorder}), where
${\varepsilon}_i$ is randomly chosen at each site with a uniform distribution in the range $[-W/2,W/2]$.
We assume that the lattice is deep enough so that the disorder alters only $\varepsilon_i$ but not
the strength of tunneling between neighboring sites $t$ and the on-site interatomic attractive interactions $U$.~\cite{Damski} We emphasize here that the attractive contact interaction has recently been realized in experiments with quantum gases by means of Feshbach resonances.~\cite{attr_inter} Experimentally the uniform random disorder can be simulated approximately by using a quasiperiodic disorder.~\cite{Lye}

Theoretical studies of disordered systems suffer a few difficulties. Firstly one has to simulate either small samples with a large amount of realizations of disorder, or to simulate very large samples. The required self-averaging makes sure that the system characterization is independent of one particular disorder realization. Due to the finite size of the real OLs, we concentrate in this paper on one-dimensional chains of few hundred lattice sites. Extension to 1000 lattice sites does not increase the numerical effort substantially. Secondly the interplay between strong interactions and disorder is a true challenge. To overcome these difficulties, we apply the Bethe-ansatz-based density-functional theory within the local density approximation (BALDA), which uses the exactly solvable one-dimensional Luther-Emery liquid as the reference system. Computationally, BALDA takes a few seconds to several minutes instead of a few hours to a few days when using the
quantum Monte Carlo (QMC) and density-matrix renormalization group (DMRG) for a single density profile calculation.
But unlike BALDA,  QMC and DMRG also provide an access to correlation functions and to the momentum distribution. 
In this study, the ground state properties in the presence of disorder are obtained by means of a disorder ensemble average,
$
\langle\langle {\cal O}\rangle\rangle_{\rm dis}=
\lim_{{\cal M}\rightarrow \infty}\frac{1}{\cal M}
\sum_{\alpha=1}^{\cal M}{\cal O}(\alpha)
$.
For example, the site occupation ${\cal N}_i$ is calculated as, ${\cal N}_i=\langle\langle n_i\rangle\rangle_{\rm dis}$, where $n_i$ is the ground-state site occupation defined in Eq.~(\ref{eq:closure}). The highly efficient computational methods allow us to perform a large amount of disorder realizations. Averaging over a large number of realizations is necessary as the strength of disorder increases. In this study, we take ${\cal M}=10^4$.
We find that the density profiles are stable against a further increase of disorder realizations.

\section{Lattice density-functional theory and Thomas-Fermi approximation}
\label{section:theory}

\subsubsection{Lattice density-functional theory}

The so-called site-occupation functional theory (SOFT)~\cite{soft,Gao_PRBlong,Schenk}
is a powerful tool to calculate the ground-state (GS) properties of an inhomogeneous lattice Hamiltonian. In our case the inhomogeneity is caused by harmonic potential and disorder. Within SOFT the exact GS site occupation,
\begin{equation}\label{eq:closure}
n_i=2\sum_{\alpha=1}^{{\rm occ.}}\left|\varphi^{(\alpha)}_i \right|^2\,,
\end{equation}
can be obtained by solving self-consistently the lattice Kohn-Sham (KS) equations
\begin{equation}\label{eq:sks}
\sum_{j=1}^{N_s}[-t_{i,j}+v^{\rm \scriptscriptstyle KS}_i \delta_{ij}]\varphi^{(\alpha)}_j
=\epsilon^{(\alpha)}\varphi^{(\alpha)}_i.
\end{equation}
Here, the effective KS potential is given by
$v^{\rm \scriptscriptstyle KS}_i
=U n_i/2+v^{\rm xc}_i+V_{2}(i-N_s/2)^2+\varepsilon_i$.
The sum over $\alpha$ runs over all the occupied orbitals, and factor 2 in Eq.~(\ref {eq:closure}) is due to spin degeneracies. The first term in the Kohn-Sham potential is the Hartree mean-field contribution,
while $v^{\rm \scriptscriptstyle xc}_i=\delta {E}_{\rm xc}[n]/\delta n_i|_{\rm \scriptscriptstyle GS}$
is the exchange-correlation (xc) potential, defined as the derivative of the xc energy
${E}_{\rm xc}[n]$ evaluated at the GS site occupation [the interested reader is encouraged to consult
Ref.~\onlinecite{Gao_PRBlong}, Appendix A for a derivation of Eq.~(\ref{eq:sks})].

The total GS energy of the system is given by
\begin{eqnarray}\label{eq:gs_energy}
{E}[n]
=\sum_\alpha \epsilon^{(\alpha)}-\sum_i v^{\rm xc}_i n_i
-\sum_i U n^2_i/4+{E}_{\rm xc}[n]\,.
\end{eqnarray}
In the actual calculation, ${E}_{\rm xc}[n]$ has to be approximated.
In this work we employ a Bethe-ansatz-based local density approximation for the xc potential,
\begin{equation}\label{eq:balda}
\left.v^{\rm xc}_{i}\right|_{\rm BALDA}
=\left. v^{\rm hom}_{\rm xc}(n,u)\right|_{n\rightarrow n_i}\,,
\end{equation}
where, in analogy with {\it ab initio} DFT, the xc potential $v^{\rm hom}_{\rm xc}(n,u)$ of the
1D homogeneous Hubbard model is defined by
\begin{equation}\label{eq:vxchom}
v^{\rm hom}_{\rm xc}(n,u)=\frac{\partial}{\partial n}\left[\epsilon_{\rm \scriptscriptstyle GS}(n,u)
-\epsilon_{\rm \scriptscriptstyle GS}(n,0)-\frac{U}{4}n^2\right]\,.
\end{equation}
Here, $\epsilon_{\rm \scriptscriptstyle GS}(n,u)$ is the GS energy per site of the 1D homogeneous system as a function of filling $n\equiv N_f/N_s$
and interaction strength $u \equiv -|U|/t$. Thus, within the local density approximation scheme proposed in Eq.~(\ref{eq:balda}), the only necessary input is the xc potential of 1D homogeneous Hubbard model, which can be found numerically from the Bethe-ansatz equations.~\cite{lieb_wu} BALDA has been shown to provide a very good account of the GS properties of 1D inhomogeneous lattice systems.~\cite{Gao_PRBlong,Gao_PRL}
For the attractive Hubbard model under harmonic confinement, the agreement between the ${\rm BALDA}$ and ${\rm DMRG}$ is excellent in certain range of parameters.~\cite{Gao_PRL} For large $u$ and/or small values of $V_2/t$, the ${\rm BALDA}$ scheme overestimates the amplitude of the atomic density waves (ADWs) and thus improved exchange-correlation functionals are demanded.~\cite{Schenk} In this paper, we choose interaction strength and confining potential within the range of values where BALDA works well compared to  DMRG. In the strong attractive regime of Eq.~({\ref{eq:hubbard}}), a flat region of doubly-occupied sites emerges at the trap center, manifesting a state of tightly bound spin-singlet dimers. In this case, we resort to Thomas-Fermi approximation, which gives an overall shape of the density site occupation and almost the same flat region of doubly-occupied sites emerging at the trap center but misses the density oscillation at the edges and atom tunneling beyond the Thomas-Fermi radius.

\subsubsection{Thomas-Fermi approximation}

In the strong attractive limit, we calculate the GS properties of $\hat{H}$ in Eq.~(\ref{eq:hubbard-disorder}) within LDA through calculating the chemical potential of the system from the local equilibrium condition,
\begin{equation}
\mu=\left.\mu_{\rm hom}(n,u)\right|_{n\rightarrow n_i}+V_2 (i-N_{s}/2)^2 + \varepsilon_i \,,
\end{equation}
derived from the direct minimization of total energy functionals,
\begin{eqnarray}
E[n]&=&\left.\sum_i \epsilon_{\rm \scriptscriptstyle GS}(n,u)\right|_{n\rightarrow n_i}
\\
&&+\sum_i V_2 (i-N_{s}/2)^2 n_i+\sum_i \varepsilon_i\,n_i\,.\nonumber
\end{eqnarray}
The particle number is determined by the  normalization condition $N_f=\sum^{N_s}_{i=1} n_i$.
Here, $\mu_{\rm hom}(n,u)$ is the chemical potential of the homogeneous system calculated from an appropriate set of Bethe-ansatz equations. This approach is termed as the Thomas-Fermi approximation (TFA),~\cite{Gao_PRBlong}
equivalent to using an LDA also for the noninteracting kinetic energy functional in the DFT scheme. The TFA
takes a non-interacting approximation for the kinetic energy but considers the exchange-correlation energy of the corresponding 1D homogeneous system, which is incorporated in $\mu_{\rm hom}(n,u)$. The performance of TFA is compared with the DMRG in the next section for the case of large $|u|$.

\section{Numerical results and discussion}

Without harmonic confinement and disorder ($V_2=0$, $W=0$), the model described by $\hat{H}$ has been exactly solved by Lieb and Wu using  the Bethe-ansatz. For attractive interactions it belongs to the Luther-Emery liquid universality class. Luther-Emery liquids are Luttinger liquids
exhibiting gapless density excitations and gapped spin excitations. Without disorder ($W=0$), this system exhibits for weak to intermediate attractive interaction compound phases characterized by the coexistence of spin pairing and atomic-density waves. ~\cite{Gao_PRL} The appearance of atomic-density waves in 1D Fermi gases with attractive interactions under confinement is due to the in-phase Friedel oscillations of spin-resolved ground-state density profiles. For strong atom-atom attractive interactions, a state of tightly bound spin-singlet dimers of doubly occupied sites appears at the center of the trap.~\cite{Gao_PRL}

\begin{figure}
\begin{center}
\includegraphics[width=0.8\linewidth]{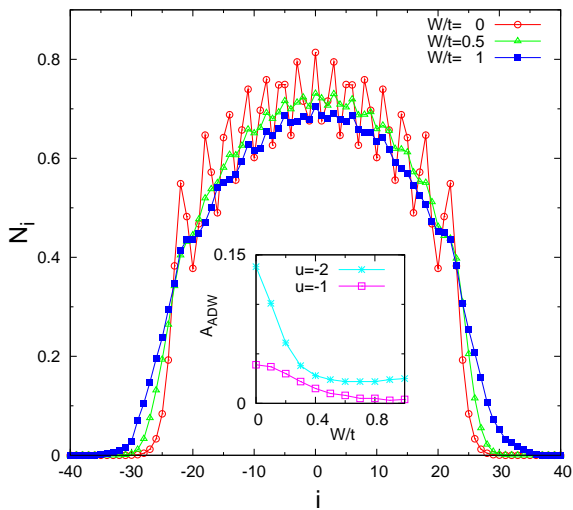}
\caption{(color online) Site occupation ${\cal N}_i$ as a function of site position $i$ for a system of
$N_f=30$, $N_{s}=100$ and $u=-2$, $V_2/t=1.0\times 10^{-3}$.
The inset shows the amplitude of the ADW, $A_{ADW}$, as a function of the strength of the disorder $W/t$ for
different interaction strengths of $u=-2$ and $u=-1$.
\label{fig:one}}
\end{center}
\end{figure}

Here we focus on the interplay between attractive interactions and disorder. For the weak-to-intermediate attractive interaction, we perform full DFT calculations for disorder realizations up to $10^4$ disorder ensemble average. In Fig.~\ref{fig:one}, we show the GS site occupation for $N_f=30$ atoms in a lattice of $N_s=100$ sites, inside a trap with $V_2/t=10^{-3}$. Without disorder, the atomic density wave is clearly seen as a consequence of Luther-Emery pairing and the formation of stable spin-singlet dimers between different pseudospins. For $W=0$, the density profile calculated from BALDA has been compared with the one from DMRG and both are in a very good agreement.~\cite{Gao_PRL} At $W=0.5$ the ADW is largely suppressed by the disorder. For large values of $|u|$, the amplitude of ADWs does not scale to zero when spin-singlet dimers occupy the deepest valleys in the disorder landscape due to the attractive nature of interaction. As a result, there is small oscillations in the density profile even in the presence of strong disorder, which is contrary to the repulsive case. In the inset, the amplitude of the ADW ($A_{ADW}$), which is well defined close to the trap center, is shown for increasing disorder strength for two different values of $|u|$, $u=-1$ and $u=-2$. The density distribution described here can be directly observed with the technique used by F\"{o}lling et al.~\cite{Foelling} They have observed the density distribution of a superfluid and a Mott insulator plateau for bosonic quantum gases in an optical lattice. We expect that the site occupation of fermionic systems will be directly detected in the near-future.

\begin{figure}
\begin{center}
\includegraphics[width=0.8\linewidth]{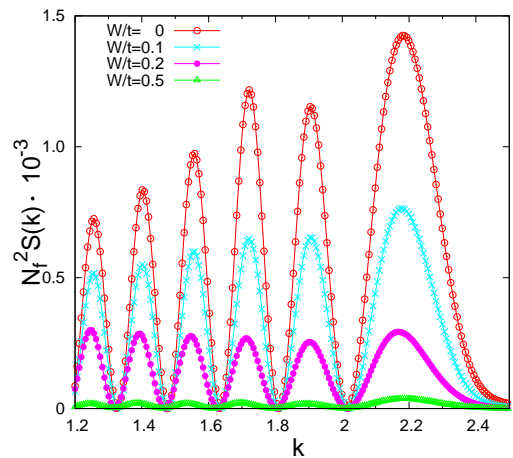}
\caption{(color online) Fraunhofer structure factor $S(q)$ for different disorder strengths in a system of
$N_f=30$, $N_{s}=100$, $u=-2$, and in the presence of a harmonic potential $V_2/t=1.0\times 10^{-3}$.
\label{fig:two}}
\end{center}
\end{figure}

The effects of disorder on the ADWs can be studied through elastic light-scattering diffraction experiments~\cite{Gao_PRL,Monica} by means of the Fraunhofer structure factor,
\[
S(k)=\frac{1}{N_f^2} \left|\sum_j \exp(-ikj)n_j\right|^2\,.
\]
The Fraunhofer structure factor has a maximum around the wave number $k=k_{ADW}\approx \pi \tilde{n}$, with $\tilde{n}$ being the
average density in the bulk of the trap. Fig.~\ref{fig:two} shows the structure factor for $N_f=30$, $N_{s}=100$, $u=-2$ and in the presence of a harmonic potential $V_2/t=1.0\times 10^{-3}$. Results for three cases of $W$ are compared with the result for $W=0$. In the system of $W=0$, $k_{ADW}\approx 0.22$, in which the Fraunhofer structure factor has a maximum. We can see how the ADW is suppressed by the disorder and thus $S(k)$ is greatly suppressed accordingly. For $W=0.5$, $S(k)$ becomes almost zero.

\begin{figure}
\begin{center}
\includegraphics[width=0.8\linewidth]{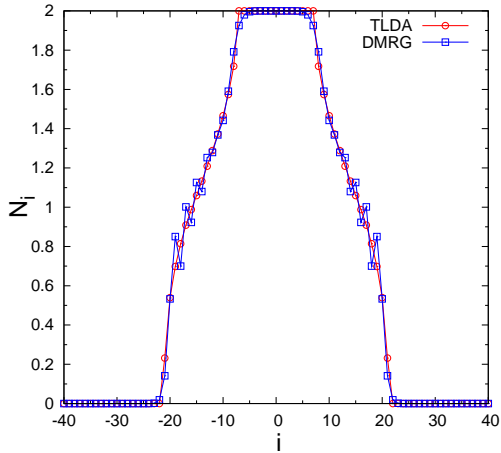}
\caption{(color online) DMRG results (open squares) for $N_f=30$, $N_{s}=200$, $V_2/t=1.0\times 10^{-3}$, and $u=-20$ in the clean limit
are compared with TFA data (open circles). The thin solid lines are just a guide for the eye.
\label{fig:three}}
\end{center}
\end{figure}

\begin{figure}
\begin{center}
\includegraphics[width=0.8\linewidth]{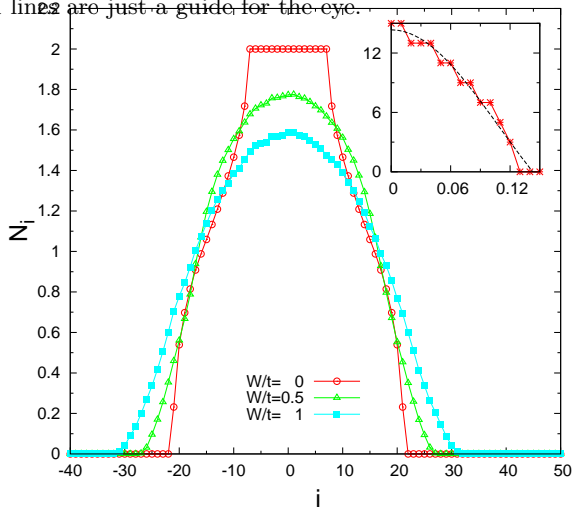}
\caption{(color online) Site occupation ${\cal N}_i$ as a function of site position $i$ for a system with
$N_f=30$ fermions in $N_{s}=200$ lattice sites, and in the presence of a harmonic potential $V_2/t=1.0\times 10^{-3}$ and strong attractive interaction of $u=-20$.
In the inset, the number of consecutive sites $N_{dimers}$ at which $|{\cal N}_i-2|\le 10^{-5}$
is shown against the strength of the disorder $W/t$.
The dashed line is the fitted exponential function: $N_{dimers}=-10.438+24.775 e^{-42.327 (W/t)^2}$.
\label{fig:four}}
\end{center}
\end{figure}

We further analyze the effect of disorder on the tightly bound spin-singlet dimers at a strong coupling. In Fig.~\ref{fig:three}, we compare the density profile calculated from TFA with the one calculated by using DMRG. As it is clear, TFA gives us a very good overall shape for the strong attractive interaction. The bound insulating region of tightly bound spin-singlet dimers at the center  of the trap is quite unstable against disorder. This region shrinks quickly with an increasing disorder amplitude and vanishes for a very small disorder value. The number of consecutive sites in the band-insulating range ($N_{\rm dimer}$) is plotted as a function of the amplitude of disorder in the inset of Fig.~\ref{fig:four}, which shows that $N_{\rm dimer}=0$ at $W=0.124$.

\begin{figure}
\begin{center}
\includegraphics[width=0.8\linewidth]{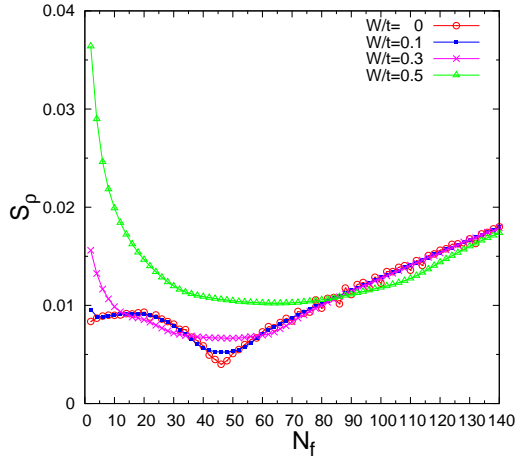}
\caption{(color online) Thermodynamic stiffness $S_{\rho}$ (in units of $t$) as a function of $N_f$ for
$u=-20$, $V_2/t=10^{-3}$ and $N_s=200$ lattice sites.
\label{fig:five}}
\end{center}
\end{figure}

In Fig.~\ref{fig:five}, we illustrate the effect of disorder on the stiffness. The stiffness is defined as, $S_\rho=\langle\langle \delta \mu/\delta {N_f}\rangle\rangle_{\rm dis}$, which is the inverse of the global compressibility, and gives us precious information on the phases of the system.
We would like to mention here, that the different quantum phases can be determined by time-of-flight experiments or by measuring noise correlations etc..~\cite{Greiner,noise}

In the absence of disorder the ground state shows two qualitatively different
phases separated by one nonanalyticity point: one phase characterized by the Luther-Emery liquid and the ADW,
while the other with a band-insulator core of tightly bound spin-singlet dimers surrounded by Luther-Emery layers.
The increase in stiffness is related to the incompressible nature of the insulating phase.

The disorder has two main effects. Firstly it leads to an anomalous behavior in the stiffness at low density,
which is similar to what has been found in the 1D harmonically trapped Hubbard model of repulsive interactions.~\cite{gao_PRB_repulsive} This phenomenon could be explained by the concept of density percolation. The atoms tends to occupy the deepest valleys in the disorder landscape which makes the high-density regions disconnected in the low density limit. In other words, the system becomes more stiff. We conclude that the low-density anomaly is robust against the nature of the interaction. Secondly, the disorder leads to the smearing of the nonanalytic point in the clean limit, signaling that the band-insulator core of tightly bound spin-singlet dimers is destroyed.

In summary, on the basis of the Bethe-ansatz based local density-functional theory and the Thomas-Fermi approximation, we have shown how disorder affects atomic density waves and spin-singlet dimers of attractively interacting Fermi gases in 1D lattices under harmonic trapping potential. We have demonstrated that atomic density waves are sensitive to the uniformly distributed disorder. The tightly bound spin-singlet dimers in the center of the trap are quite unstable against the disorder. An anomalous increase of the stiffness persists at low density from quenching of percolation, which has also been found in repulsive systems. We have shown that the nonanalyticity point in the stiffness signals the phase transition from the compressible liquid phase to the incompressible insulating phase with spin-singlet dimers in the center. It is found that very weak disorder leads to the smoothing of the nonanalyticity point in the stiffness, thus suppresses the spin-singlet pairs.

\begin{acknowledgments}
This work was supported by NSF of China under grand No. 10704066 and Qianjiang River Fellow Fund 2008R10029. The author gratefully acknowledges many useful discussions with M. Polini,  S. Montangero, F. Moura, S.H. Abedinpour and P. Yu. The DMRG code of the "Powder with Power" project (http://www.qti.sns.it) has been used.
\end{acknowledgments}

\end{document}